\documentclass[twocolumn]{aastex61}

\newcommand\aastex{AAS\TeX}

\newcommand{\msun}{\hbox{M$_{\odot}$}}
\newcommand{\kms}{\hbox{km s$^{-1}$}}

\newcommand{\mbh}{$M_\bullet$}

\received{}
\revised{}
\accepted{\today}
\shorttitle{\aastex\ Exploring the limits of AGN feedback}
\shortauthors{MM}
\begin{document}

\title{Exploring the limits of AGN feedback: \\black holes and the star formation histories of low-mass galaxies}

\correspondingauthor{I. Mart\'in-Navarro}
\email{imartinn@ucsc.edu, marmezcua.astro@gmail.com}

\author[0000-0003-4266-5580]{I. Mart\'in-Navarro}
\affil{University of California Observatories, 1156 High Street, Santa Cruz, CA 95064, USA}
\affil{Max-Planck Institut f\"ur Astronomie, Konigstuhl 17, D-69117 Heidelberg, Germany}

\author{M. Mezcua}
\affiliation{Institute of Space Sciences (ICE, CSIC​)​, Campus UAB, Carrer de Can Magrans, 08193, Barcelona, Spain}
\affiliation{Institut d'Estudis Espacials de Catalunya (IEEC), C/ Gran Capit\`{a}, 08034 Barcelona, Spain}



\begin{abstract}
Energy feedback, either from active galactic nuclei (AGN) or from supernovae, is required to understand galaxy formation within a $\Lambda$-Cold Dark Matter cosmology. We study a sample of 127 low-mass galaxies, comparing their stellar populations properties to the mass of the central supermassive black hole, in order to investigate the effect of AGN feedback. We find a loose coupling between star formation history and black hole mass, which seems to suggest that AGN activity does not dominate baryonic cooling in low-mass galaxies. We also find that a break in the \mbh-$\sigma$ relation marks a transitional stellar mass, M$_\mathrm{trans}=3.4\pm2.1 \times 10^{10}$ \msun, remarkably similar to M$_\star$. Our results are in agreement with a bi-modal star formation process where the AGN-dominated feedback of high-mass galaxies transitions towards a supernovae-driven regime in low-mass systems, as suggested by numerical simulations.
\end{abstract}




\section{Introduction} 
The finding of tight correlations between the mass of supermassive black holes (SMBHs) at the center of massive galaxies and some of the host galaxy properties such as stellar velocity dispersion \citep[the \mbh-$\sigma$ relation; e.g.][see \citealt{Kormendy13} for a review]{Gebhardt00,Ferrarese00}

strongly suggest that the evolution of massive galaxies must be linked to that of their central SMBHs. Feedback from the active galactic nucleus (AGN) onto the galaxy is thought to regulate the formation of stars (and thus stellar growth) and is a necessary ingredient in cosmological simulations in order to solve the ``over-cooling'' problem (i.e. the too high star formation rate -SFR- predicted in the absence of feedback) and correctly reproduce the observed galaxy luminosity function, which breaks at L$_\star$ \citep[e.g.][]{Croton06,Bower17,Choi17}. Observational evidence for such a coupling between SMBH activity and galaxy star formation comes from the finding in galaxy clusters of large-scale X-ray activities inflated by the jet of the central SMBH \citep[e.g.][]{Fabian00,McNamara00,HL12} and, more recently, from the dependence of BH mass with SFR found for nearby massive galaxies \citep{MN16,MN17}.

Low-mass galaxies, at the faint end of the luminosity function, exhibit bluer colors and have higher levels of star formation activity than red massive galaxies in which star formation is almost undetectable \citep[e.g.][]{Gallazzi05,thomas05}. In these low-mass galaxies the outflows from young stars and supernovae (SNe) are thought to deplete the central BH from its gas reservoir, hampering BH growth and thus the impact of AGN feedback \citep[e.g.][]{Dubois15,Bower17}. The transition from the AGN feedback regime that dominates massive galaxies to the SN feedback governing low-mass galaxies is predicted by simulations to be driven by a transition in galaxy properties at L$_\star$ or at a mass M$_\star$ of $\sim3 \times 10^{10}$ M$_{\odot}$ and to occur already in the early Universe \citep[e.g.][]{Crain15,Bower17}. Compelling observational evidence that this SN-regulated BH growth dominates over AGN feedback in dwarf galaxies is however so far lacking. The increasing number of accreting BHs being recently found in dwarf galaxies has allowed us to change this. 

The finding that SMBHs were already in place at z$>$6 suggests that they had to grow from seed BHs of $10^{2} <$ \mbh $< 10^{6}$ M$_{\odot}$ in order to be able to reach a mass of more than 10$^{9}$ M$_{\odot}$ in such a short time \citep[e.g.][]{Volonteri10}. In the nearby Universe, such seed or intermediate-mass BHs (IMBHs) should be found in dwarf galaxies resembling the first galaxies formed at z$>$6, with low mass and low metallicity. Optical searches for IMBHs in low-mass galaxies based on the assumption that the gas around the active BH is virialized has yielded several hundreds of candidates (e.g. \citealt{Greene04,Greene07,Reines13}), in $\sim$25\% of which the AGN nature has been reinforced by the detection of X-ray emission (e.g. \citealt{Desroches09,Dong12,Baldassare15,Baldassare17}). Direct X-ray searches have yielded the identification of a few more tens of AGN in dwarf galaxies (e.g. \citealt{Reines14,Pardo16,Mezcua16,Mezcua18}; see \citealt{Mezcua17} for a review).

In this Letter we use a large sample of more than 100 low-mass galaxies with signs of AGN activity indicated by the presence of broad emission lines in their optical Sloan Digital Sky Survey (SDSS) spectrum and study any dependence of the BH mass with star formation activity. The absence of a clear BH-SFR connection, unlike that found for massive galaxies \citep{MN17}, implies that AGN feedback is not significant in these galaxies but that SN feedback must drive their growth. This is in agreement with the predictions of cosmological simulations and with the plateau observed in the low-mass regime of the \mbh-$\sigma$ correlation.

\section{Sample selection}
Our study includes 127 low-mass Seyfert 1 galaxies with measured stellar velocity dispersions and BH masses. This compilation is based on the combination of the samples presented in \citet{Xiao11} and in \citet{Woo15}, both drawn from the SDSS. \citet{Xiao11} based their analysis on the original sample of \citet{Greene07}, which selected all broad-line active galaxies with \mbh$<2\times10^6$ M$_{\odot}$ in the SDSS DR4. \citet{Woo15} used the SDSS DR7 to identify a sample of 464 narrow-line Seyfert 1 galaxies. 

Stellar velocity dispersions among \citet{Xiao11} and in \citet{Woo15} samples were measured differently. \citet{Xiao11} used spectroscopic data obtained with the Keck and the Clay telescopes. These observations were complemented with the velocity dispersions measurements of \citet{Barth05}. For the \citet{Woo15} sample, stellar velocity dispersions could be directly measured from the SDSS spectra for 63 objects, combining different spectral regions in order to minimize systematics in the analysis. BH masses were estimated under the assumption that the gas is virialized and following the empirically estimated expression:
\begin{equation}
 \log \mathrm{M}_\bullet = \alpha + \beta \bigg(\frac{L_{\mathrm{H}_\alpha}}{10^{42} \ \mathrm{erg \ s}^{-1}}\bigg) + \gamma \bigg(\frac{\mathrm{FWHM}_{\mathrm{H}_\alpha}}{10^{3} \ \mathrm{km \ s}^{-1}}\bigg) 
\end{equation}

\noindent where $L_{\mathrm{H}_\alpha}$ and $\mathrm{FWHM}_{\mathrm{H}_\alpha}$ are the luminosity and the full width at half maximum (FWHM) of the broad-component of the H$_\alpha$ line, and $\alpha$, $\beta$, and $\gamma$ are calibration constants depending on modeling assumptions.  \citet{Xiao11}  adopted $\alpha=6.40$, $\beta=0.45$, $\gamma=2.06$, while \citet{Woo15} assumed $\alpha=6.544$, $\beta=0.46$, $\gamma=2.06$. The typical systematic uncertainty of BH mass measurements using this spectroscopic technique is of $\sim$0.3 dex.

\subsection{The (local) \mbh-$\sigma$ relation}
Following \citet{MN16}, we investigate the interplay between BH activity and star formation using the \mbh-$\sigma$ relation as a metric. At a given $\sigma$, galaxies above/below the best-fitting relation have a heavier/lighter BH than the average population. Since the energy released scales up with the mass of the BH \citep{Crain15,Sijacki15}, galaxies with more massive BHs, e.g. above the relation, would have experienced a more intense feedback. Therefore, studying the stellar population properties across the \mbh-$\sigma$ relation probes how star formation depends on the energy input from the central BH. The universality of the \mbh-$\sigma$ relation allows us to isolate the effect of BH feedback.

\begin{figure}
\begin{center}
\includegraphics[width=8cm]{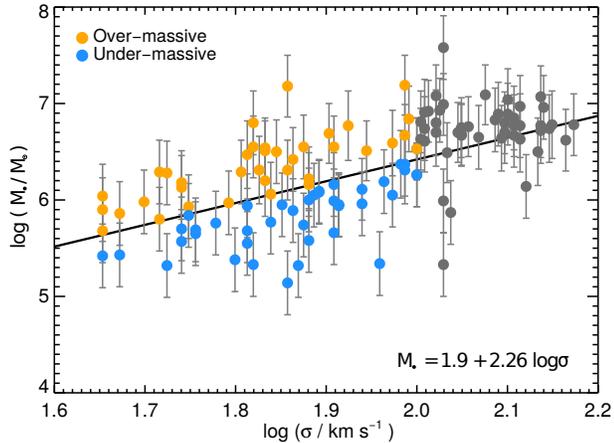}
\end{center}
\caption{\mbh-$\sigma$ relation for Seyfert 1 galaxies. We restrict our stellar population analysis to over-massive (orange) and under-massive (blue) BH galaxies with $\sigma<100$ km s$^{-1}$. Black line indicates the best-fitting relation for these low-$\sigma$ galaxies. Error bars include both systematic and observational uncertainties.}
\label{fig:msigma}
\end{figure}

In Fig.~\ref{fig:msigma} we show the \mbh-$\sigma$ relation defined by our sample. As described above, L$_\star$ galaxies mark a transition between SN-regulated star formation in low-mass systems, and the AGN-dominated regime of massive galaxies. Since the main motivation of this work is to constrain the effect of AGN feedback in a mass range where it is sub-dominant, we restrict our analysis to galaxies with $\sigma<100$ km s$^{-1}$ (colored symbols in Fig.~\ref{fig:msigma}). Using the Bayesian linear regression scheme described in \citet{Kelly07}, we find that our sample of low-mas ($\sigma<100$ km s$^{-1}$) galaxies follow a best-fitting relation given by

\begin{equation} \label{eq:rela}
 \log M_\bullet = 1.9 + 2.26 \log \sigma
\end{equation}

\noindent Using this relation, shown as a solid line in Fig.~\ref{fig:msigma}, we sub-divide our sample into over-massive and under-massive BH galaxies. Over-massive BH galaxies are those objects above the relation (orange symbols), where the AGN effect is expected to be maximum, and under-massive BH galaxies (in blue) are those below the relation. The average velocity dispersion is $\sigma=69.3\pm$2.6 km s$^{-1}$  and $\sigma=73.8\pm$2.3 km s$^{-1}$ for over-massive and under-massive BH galaxies, respectively. This translates into a typical stellar mass of M$_\star \sim 0.9\times10^{10}$ \msun. By construction, over-massive BH galaxies host heavier BHs than under-massive BH objects ($\log$\mbh$^\mathrm{OM}=6.31\pm0.05$ \msun \ and  $\log$\mbh$^\mathrm{UM}=5.89\pm0.05$ \msun). Note that Eq.~\ref{eq:rela} should not be taken as an absolute reference, but rather as a local measurement. The exact behavior of the \mbh-$\sigma$ relation for low-mass galaxies and how it relates to higher-mass objects is still an open question \citep[e.g.][]{Wyithe06,Graham09,Xiao11,Mezcua17}, and will be discussed in detail in \S~\ref{sec:results}. For the purposes of this work, our best-fitting relation is fundamentally a {\it local} ruler to quantify how heavy/light a given BH is compared to the average population, at fixed stellar velocity dispersion.

\section{Data and stellar populations analysis}
We have based our stellar population analysis on spectra from the SDSS DR12. Unfortunately, the low signal-to-noise of the individual spectra is not high enough to obtain reliable star formation histories (SFHs). To meet the signal-to-noise requirements, we perform a Voronoi binning of the \mbh-$\sigma$ plane \citep{Cappellari03}. In order to avoid confusion issues around the best-fitting relation, we first divide our sample into over-massive and under-massive BH galaxies, and we latter carry out the two-dimensional binning (i.e., \mbh \ and $\sigma$) considering the two sub-samples separately. We impose a minimum signal-to-noise of 35, which leads to $\sim5$ bins within both over-massive and under-massive BH galaxies. Notice that having multiple representative spectra above and below the best-fitting \mbh-$\sigma$ relations minimizes the effect of systematics in the stellar population analysis.

\begin{figure}
\begin{center}
\includegraphics[width=8cm]{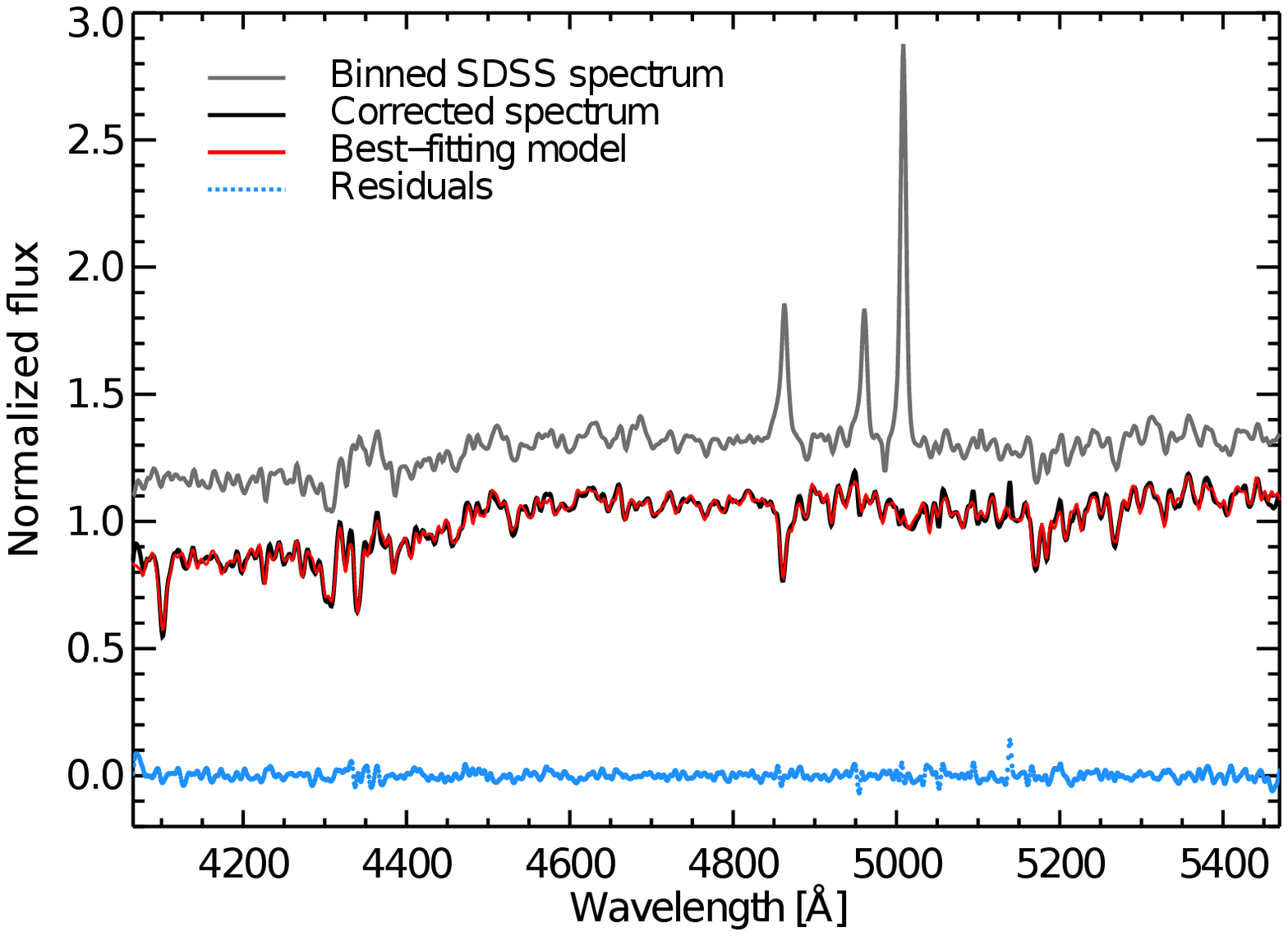}
\end{center}
\caption{Spectroscopic data. We show in gray a typical Voronoi-binned SDSS spectrum, and in black the same spectrum after being corrected for non-stellar features (nebular emission plus FeII and AGN continuum). The best-fitting stellar populations model is shown in red, along with the percent-level residuals in blue.}
\label{fig:bfit}
\end{figure}

After the Voronoi binning, we proceed to generate the combined spectra. We use the SDSS redshift measurements to correct for the radial velocity of each individual galaxy. We then normalize the spectra by the median flux over the 4000$<\lambda<5000$ \AA \ wavelength range. We also homogenize the spectral resolution of the SDSS data by smoothing the data to a common resolution of FWHM$=5$\AA, using a gaussian kernel. Although the wavelength dependence of the SDSS resolution is sometimes ignored, we make sure in this way that we do not introduce artifacts on the stellar population analysis. After applying the redshift and spectral resolution corrections, we co-add individual spectra following our Voronoi binning tessellations. In Fig.~\ref{fig:bfit}, the gray line shows one of these combined SDSS spectra. These are dominated by strong nebular emission lines, in addition to the Balmer H$_\delta$, H$_\gamma$, and H$_\beta$ lines, which we subtract before attempting the stellar population analysis. To estimate the best-fitting stellar template we make use of the Penalized Pixel-Fitting \citep[][pPXF]{ppxf1}. We then use the residuals from this first fitting to individually model each emission line as the sum of two gaussians. There is no physical motivation behind a two-component fitting, but it provides a reasonably good fit to the observed spectra (see Fig.~\ref{fig:bfit}). Notice also that we are not interested in the properties of the emission lines.

In addition to the nebular emission, we also need to remove from our spectra the non-stellar FeII emission and the AGN continuum. This time, we feed pPXF with the FeII templates of \citet{feii}, including also a featureless (additive) continuum to model the AGN contribution. The black solid line in Fig.~\ref{fig:bfit} shows one of our corrected SDSS spectra, after accounting for all the emission features (nebular + FeII + AGN continuum). 

\subsection{Stellar populations}
To measure the stellar populations properties, and in particular the SFHs among our sample, we use the STEllar Content and Kinematics via Maximum A Posteriori likelihood (STECKMAP) code \citep{Ocvirk06b}. STECKMAP is a Bayesian algorithm to find the best-fitting linear combination of single stellar population models describing a given spectrum. It non-parametrically reconstructs SFHs and has been extensively tested \citep[e.g.][]{Pat11,RL15}. We feed STECKMAP with the MILES stellar population synthesis models \citep{Vazdekis10}. These models cover a range in metallicity from $-1.3$ to $+-0.22$ dex, and from 67 Myr to 17 Gyr in age. We use the base MILES models, which follow the solar neighborhood abundance pattern, i.e., they are $\alpha$-enhanced ([$\alpha$/Fe] $\sim +0.4$) for metal-poor populations, but becoming [$\alpha$/Fe] $\sim 0.0$ for solar and super-solar metallicities. STECKMAP allows for a simultaneous analysis of the kinematical and stellar population properties. The solid red line in Fig.~\ref{fig:bfit} shows the best-fitting model given the (black solid line) spectrum. Residuals (in blue) are of the order of a few percent, with no significant structure. This low level of scatter in the residuals, in particular around the prominent emission lines, suggests that the correction from AGN contamination is accurate enough for a robust stellar population analysis.

\section{Results}
\label{sec:results}
\begin{figure}
\begin{center}
\includegraphics[width=8cm]{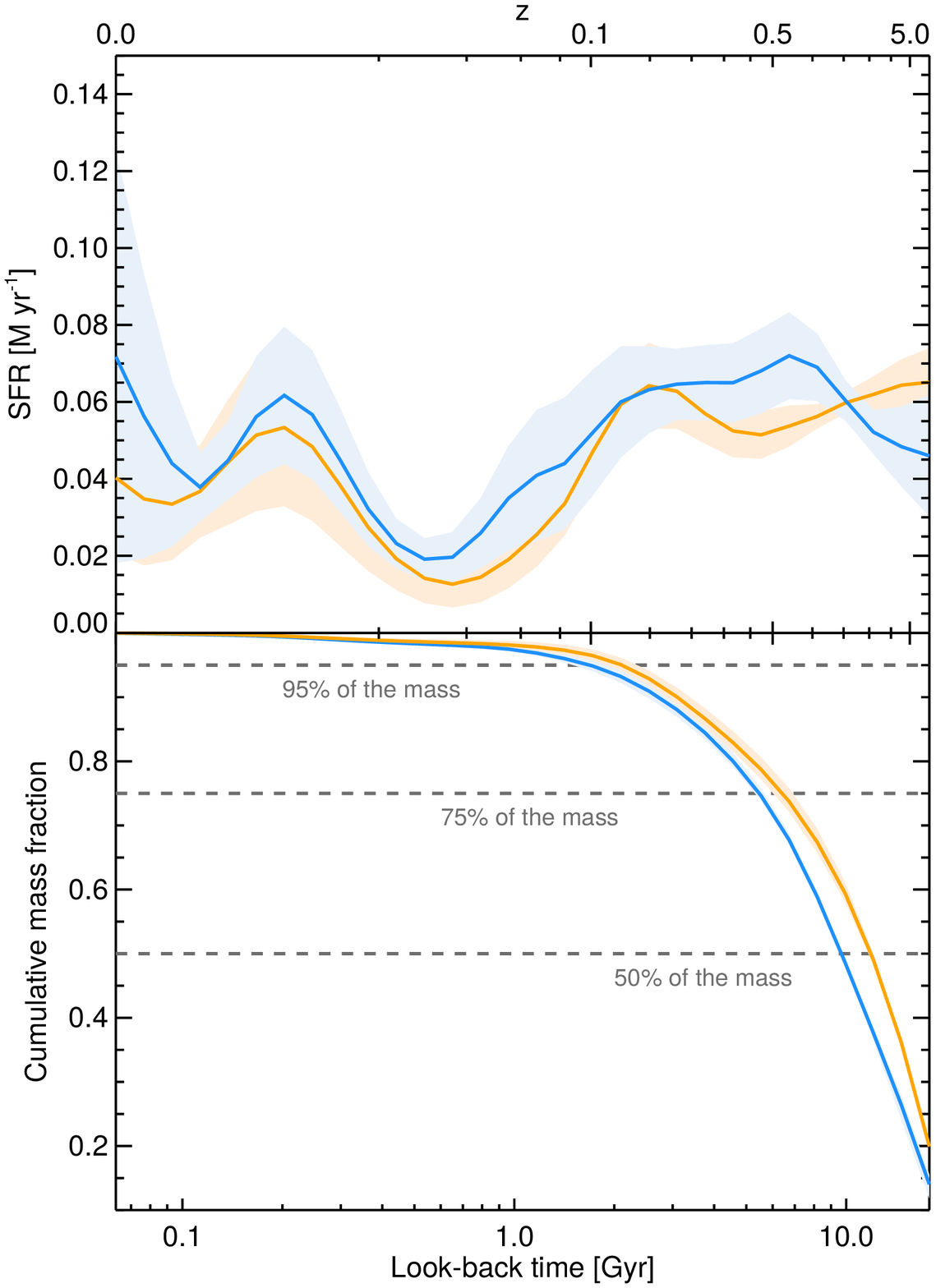}
\end{center}
\caption{SFR (top) and cumulative stellar mass function (bottom) as a function of look-back time for under-massive (blue) and over-massive (orange) BH galaxies. Shaded regions correspond to the 1$\sigma$ uncertainty of the mean value. The differences between both types of objects are weak, negligible if the b/a is taken into account. Dashed horizontal lines in the bottom panel indicate when 50\%, 75\% and 95\% of the galaxy mass is reached, respectively.}
\label{fig:sfh}
\end{figure}

The average SFHs and cumulative stellar mass distributions for over-massive and under-massive (low-$\sigma$) galaxies are shown in Fig.~\ref{fig:sfh}. Shaded regions indicate the error around the mean value. The SFR as a function of look-back time shows a rather bursty behavior at all redshifts, characteristic of low-mass objects \citep[e.g.][]{Koleva09}.
 
Over-massive and under-massive BH galaxies in the explored $\sigma$ range exhibit rather similar SFHs. The main differences appear for very old stellar populations, where the SFR in galaxies above the best-fitting \mbh-$\sigma$ relation seem to be slightly elevated compared to those galaxies below the relation. This is reflected in the mass-weighted age, which is marginally older for over-massive than for under-massive BH galaxies (5.2$\pm$0.6 and 4.3$\pm$0.4 Gyr, respectively).

We note that the broad emission lines used in our sample as a proxy for the BH mass could also originate from transient stellar phases \citep[e.g.][]{Baldassare16}. To reinforce the AGN origin of the emission lines, we repeat the analysis but using only those objects in our sample with detected nuclear X-ray emission as a signature of AGN activity (39 out of 127 dwarf galaxies; e.g.  \citealt{Desroches09,Dong12}). The results we obtain are consistent with those found for the full sample of 127 objects.

Galaxy inclination might artificially shift the position of a given object in the \mbh-$\sigma$, as the effective velocity dispersion of a face-on rotating system becomes meaningless. Given the predominantly disky nature of our sample, this effect may significantly influence our results. We test this by homogeneously deriving the b/a axis ratios of our sample based on the SDSS DR12 photometry. We perform then a bi-linear fitting of the \mbh-$\sigma$-b/a ratio plane, finding that log $M_\bullet=1.8+2.28$ log $\sigma+0.09\mathrm{b/a}$ and that galaxies with higher b/a ratios tend to have lower $\sigma$. This is in agreement with the results from \citet{Xiao11} and \citet{Woo15} and reduces even more the differences between the SFH of the over-massive and under-massive galaxies.

As an additional test, we also create a single over-massive and under-massive BH galaxy spectrum by combining all the objects above and below the relation. The recovered SFHs are also fully compatible with those obtained with the binned spectra, suggesting that Fig.~\ref{fig:sfh} actually reflects the average behavior of both types of galaxies and it is not driven by statistical flukes. Finally, we also investigate the effect of systematic uncertainties in the BH mass estimations. Considering only those objects deviating more than 0.3 dex from the \mbh-$\sigma$ relation, the differences in the SFHs remain similar to those shown by Fig.~\ref{fig:sfh}.

\section{Discussion and conclusions}
The rather similar SFHs of under-massive and over-massive BH galaxies in the low-$\sigma$ end suggest that AGN activity may not play a dominant role in regulating star formation in low-mass galaxies. Our findings offer observational support to the bi-modal paradigm of star formation within galaxies: while SN feedback is likely to regulate baryonic cooling within low-mass dark matter halos, a different, AGN-driven regime drives star formation in massive galaxies. The transition between SN- and AGN-driven modes occurs at a characteristic stellar mass M$_\star \sim3 \times 10^{10}$ M$_{\odot}$ \citep[e.g.][]{Baldry12}. 

\begin{figure*}[!ht]
\begin{center}
\includegraphics[width=15cm]{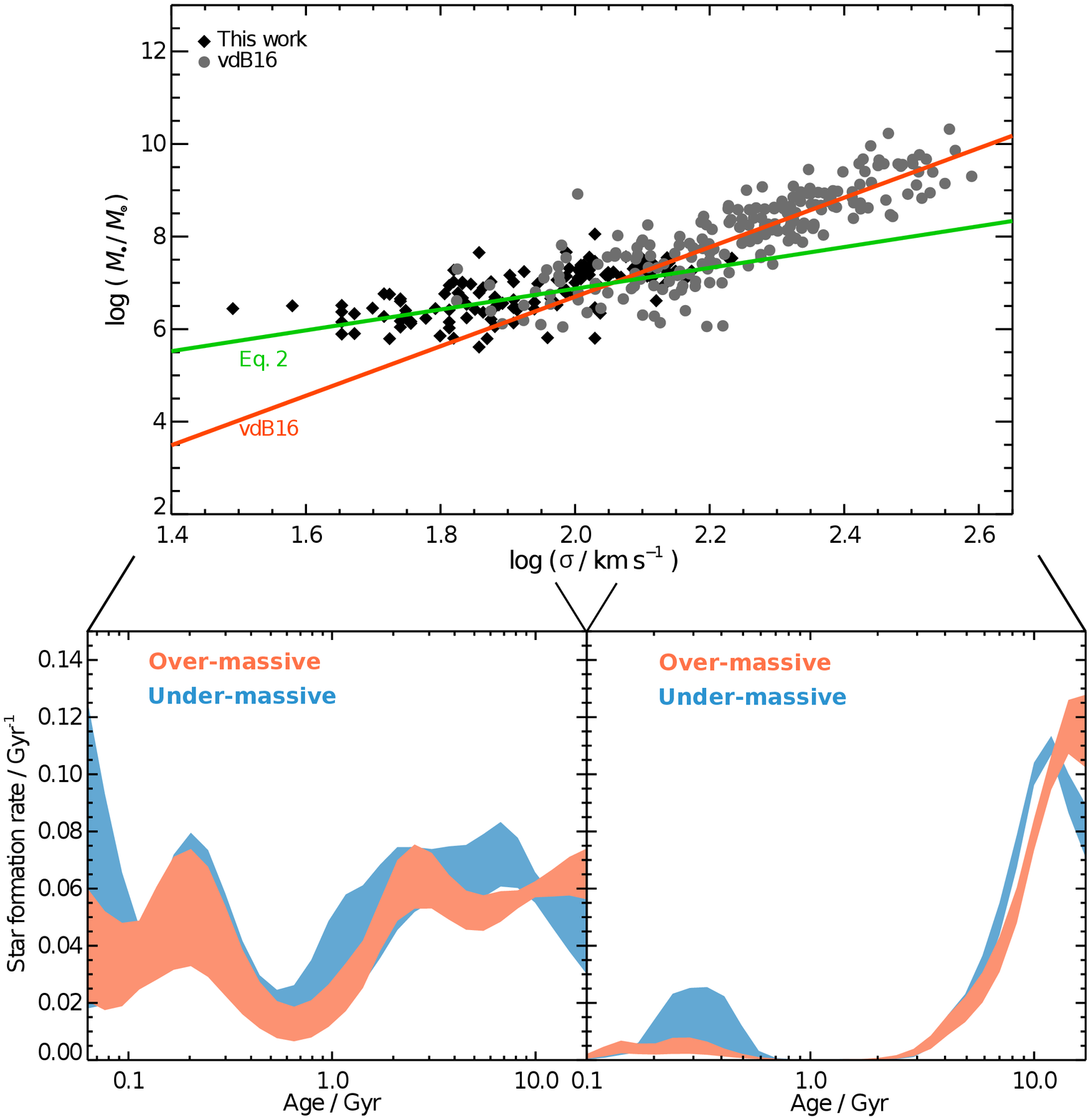}
\end{center}
\caption{Upper panel shows the \mbh-$\sigma$, combining our sample (black diamonds) with that of \citet{vdb16} for higher mass galaxies (light circles). The green line indicates our best-fitting trend for $\sigma\sim70$ km s$^{-1}$ galaxies, while the red line is the relation expected for more massive object ($\sigma\sim200$ km s$^{-1}$). A characteristic $\sigma_\mathrm{trans}\sim100$\kms marks the transition between low-mass and high-mass scaling relations, and correspond to a stellar mass of  M$_\mathrm{trans}\sim3\times10^{10}$\msun. In the bottom left and bottom right panels, we show the SFHs of over-massive (orange) and under-massive (blue) BH galaxies for low-$\sigma$ and high-$\sigma$ galaxies, respectively. We interpret the differences between these two panels as a transition process between an AGN-dominated regime for high-mass galaxies, towards a non-AGN related star formation mode for low-mass systems, probably regulated by SN feedback.}
\label{fig:combi}
\end{figure*}

To better understand this bi-modal character of star formation, we compare in Fig.~\ref{fig:combi} the SFHs of under-massive and over-massive BH galaxies along the \mbh-$\sigma$ relation (upper panel).  For high-$\sigma$ galaxies, we show the SFHs presented in \citet{MN17}. BH masses in Seyfert 1 galaxies have been scaled up by +0.4 dex to match direct BH mass measurements. It has been pointed out that the \mbh-$\sigma$ relation flattens for low-mass galaxies, with BH masses asymptotically tending towards $\log$\mbh$\sim 5$\msun \citep{Greene06,Mezcua17}, favoring a direct collapse-formation scenario \citep[e.g.][]{Volonteri10}. Such a flattening is clear from Fig.~\ref{fig:combi}, which combines our sample of 127 objects, plus 204 galaxies with direct BH mass measurements from \citet{vdb16}, being the largest comparison up to date. This flattening seems to suggest a weaker but still noticeable coupling between BHs and star formation, in agreement with our analysis. Our best-fitting \mbh-$\sigma$ relation for the 127 low-mass galaxies is shown in green, and that of \citet{vdb16} is shown in red. Note that our best-fitting relation describes the average behavior of $\sigma\sim70$\,\kms galaxies, while the sample of \citet{vdb16} probes galaxies with $\sigma\sim200$\,\kms. Interestingly, the intersection between both fits occurs at characteristic $\sigma_\mathrm{trans}=96\pm25$\kms, or equivalently, at a characteristic stellar mass of M$_\mathrm{trans}=3.4\pm2.1 \times 10^{10}$ \msun \citep{vdb16}, remarkably similar to M$_\star$. Notice that the uncertainty in M$_\mathrm{trans}$ is determined by the scaling factor applied to Seyfert 1 BH masses. The lower limit corresponds to uncorrected BH masses, and the upper limit is that shown in Fig.~\ref{fig:combi}, being this the only effect of the scaling factor on our results. 

We therefore hypothesize that our observations can be explained if the role of AGN feedback in regulating star formation becomes sub-dominant in low-$\sigma$ galaxies, as expected from numerical simulations \citep[e.g.][]{Habouzit17}. Low-$\sigma$ galaxies, with BH masses asymptotically converging towards \mbh$\sim10^5$\msun, would be the low-mass relics of the high-$z$ galaxy seeds. Since SN feedback is capable of preventing effective gas cooling towards the center of low-$\sigma$ galaxies \citep{Dubois15}, the central SMBH is not able to reach the critical mass required to regulate star formation \citep{Bower17}. The same process of {\it black hole starvation} would explain the lack of classical bulge-like morphologies in our sample \citep{Jiang11}. 

Such a scenario would also explain the flattening of the \mbh-$\sigma$ relation, as a consequence of a weaker coupling between SMBH and host galaxy. Our transition mass M$_\mathrm{trans}$, equivalent to M$_\star$, would then separate the high-mass end regime, where star formation occurs under a {\it hot accretion} mode regulated by AGN activity, and low-mass galaxies, where SNe regulate star formation. {\it Cold accretion} would be dominant in these low-mass systems, whose central BH and stellar mass have not grow enough to match the standard (high-mass end) scaling relations. However, some caveats in our analysis should be noted. First, our results consist of an upper limit for the BH effect on baryonic cooling given the uncertainties in low-$\sigma$ BH mass estimates. Second, at low galaxy stellar masses the BH occupation fraction is not 1 \citep[e.g.][]{Mezcua18} and thus our sample may not be a complete representation of the general population. Finally, we have not directly tested whether the presence of an AGN has an impact on the star formation history of the host galaxy. Such analysis is left for when the number of low-mass galaxies with dynamical BH mass measurements is large enough for such study to be performed. We note though that such impact is not expected given the radically different time-scales of AGN activity and star formation.

In summary, our results suggest that the AGN effect on baryonic cooling strongly depends on galaxy mass, being weaker in low-$\sigma$ galaxies than in high-$\sigma$ objects. Deeper optical spectra of low-$\sigma$ galaxies will be needed in the future to better constrain the stellar population properties of these objects. Moreover, larger and more complete samples of the low-mass end of the galaxy mass function are needed to overcome possible systematics related to Seyfert 1 galaxies \citep[e.g.][]{Ricci17}.

\acknowledgments
The authors thank Pau Diaz Gallifa for his constant support. I.M.N. acknowledges funding from the Marie Sk\l odowska-Curie Individual Fellowship 702607, and from grant AYA2013-48226-C3-1-P from the Spanish Ministry of Economy and Competitiveness (MINECO). M.M. acknowledges support from the Spanish Juan de la Cierva program (IJCI-2015-23944).

\bibliographystyle{aasjournal}
% \bibliography{imbh}



\end{document}